\documentclass[preprint]{aastex61}

\usepackage{graphicx}
\usepackage{float}

\received{2017 November 17}
\revised{2017 December 8}
\accepted{2017 December 16}
\submitjournal{The Astrophysical Journal Letters}

\shorttitle{SUVI Eruption Observations}
\shortauthors{Seaton \& Darnel}

\begin{document}

\title{Observations of an Eruptive Solar Flare in the Extended EUV Solar Corona}

\author[0000-0002-0494-2025]{Daniel B. Seaton}
\affiliation{Cooperative Institute for Research in Environmental Sciences, University of Colorado at Boulder, Boulder, Colorado, 80305, USA}
\affiliation{National Centers for Environmental Information, National Oceanic and Atmospheric Administration, Boulder, Colorado, 80305, USA}

\author[0000-0001-6601-9116]{Jonathan M. Darnel}
\affiliation{Cooperative Institute for Research in Environmental Sciences, University of Colorado at Boulder, Boulder, Colorado, 80305, USA}
\affiliation{National Centers for Environmental Information, National Oceanic and Atmospheric Administration, Boulder, Colorado, 80305, USA}

\correspondingauthor{Dan Seaton}
\email{daniel.seaton@colorado.edu}

%% Mark off the abstract in the ``abstract'' environment. 
\begin{abstract}
We present observations of a powerful solar eruption, accompanied by an X8.2 solar flare, from NOAA Active Region 12673 on 2017 September 10 by the Solar Ultraviolet Imager (SUVI) on the {\it GOES-16} spacecraft. SUVI is noteworthy for its relatively large field of view, which allows it to image solar phenomena to heights approaching 2~solar radii. These observations include the detection of an apparent current sheet associated with magnetic reconnection in the wake of the eruption and evidence of an extreme-ultraviolet wave at some of the largest heights ever reported. We discuss the acceleration of the nascent coronal mass ejection to approximately 2000~km/s at about 1.5~solar radii. We compare these observations with models of eruptions and eruption-related phenomena. We also describe the SUVI data and discuss how the scientific community can access SUVI observations of the event.
\end{abstract}

%% Keywords should appear after the \end{abstract} command. 
%% See the online documentation for the full list of available subject
%% keywords and the rules for their use.

%% From the front matter, we move on to the body of the paper.
%% Sections are demarcated by \section and \subsection, respectively.
%% Observe the use of the LaTeX \label
%% command after the \subsection to give a symbolic KEY to the
%% subsection for cross-referencing in a \ref command.
%% You can use LaTeX's \ref and \label commands to keep track of
%% cross-references to sections, equations, tables, and figures.
%% That way, if you change the order of any elements, LaTeX will
%% automatically renumber them.

%% We recommend that authors also use the natbib \citep
%% and \citet commands to identify citations.  The citations are
%% tied to the reference list via symbolic KEYs. The KEY corresponds
%% to the KEY in the \bibitem in the reference list below. 

\section{Introduction} \label{sec:intro}

Observations of extreme-ultraviolet (EUV) phenomena in the solar corona have largely been limited to relatively low heights due to the restricted fields of view of EUV solar imagers. These imagers include the Atmospheric Imaging Assembly \citep[AIA;][]{2012SoPh..275...17L} on board the {\it Solar Dynamics Observatory} (SDO) spacecraft and the Extreme-ultraviolet Imaging Telescope \citep[EIT;][]{Delaboud95} on the {\it Solar and Heliospheric Observatory} (SOHO) spacecraft, neither of which sees the solar corona much above 1.4~solar radii.

More recently the Sun Watcher with Active Pixels and Image Processing \citep[SWAP;][]{2013SoPh..286...43S, 2013SoPh..286...67H} on board the {\it Project for On-Board Autonomy 2} (PROBA2) spacecraft --- with a field of view of 54~arcmin on the horizontal and about 76~arcmin on the diagonal --- has demonstrated the value of EUV solar observations up to heights as large as 2.5~solar radii \citep{2013ApJ...777...72S}. However, SWAP is limited to a single passband at about 17.4~nm, which is not optimized for observations of the dynamic and high temperature phenomena associated with solar eruptions. So the advantage gained by SWAP's large field of view --- when it comes to eruptions --- has been largely restricted to phenomena that can be observed in cooler temperature ranges. These observations include features such as large post-eruptive loop systems  \citep{2015ApJ...801L...6W} and flux ropes and filaments \citep{2014SoPh..289.4545B, 2013SoPh..286..241M}.

The new Solar Ultraviolet Imager (SUVI) on NOAA's {\it GOES-16} spacecraft offers a field of view similar to SWAP's, but with observations at slightly higher resolution and wider spectral coverage, with six passbands at 9.4, 13.1, 17.1, 19.5, 28.4, and 30.4~nm. SUVI's camera also has anti-blooming circuitry that ensures a clear view of flaring regions even in exposures where the detector has saturated considerably. This circuitry has a negligible impact on SUVI's ability to observe fainter structures in the higher corona, but ensures blooming does not obscure flaring regions. SUVI is therefore particularly well-suited to making observations that span a large dynamic range like those we report here.

In this Letter we present SUVI observations of a large solar eruption accompanied by an X8.2 solar flare that occurred on the Sun's west limb at about 15:40~UT on 2017	September 10. The eruption originated from NOAA Active Region (AR) 12673, which also produced a X9.3 flare a few days earlier on September 6. Together, these events represent the most energetic outbursts of Solar Cycle 24.

Because the eruption occurred on the limb, SUVI's large field of view made possible observations of phenomena in the more extended corona that were not seen by other imagers. In this paper we focus on three key phenomena for which SUVI's observations were especially unique. First, we report on the onset of the eruption and the acceleration of the nascent coronal mass ejection (CME) associated with it. Second, we discuss how the eruption was accompanied by a narrow, bright, extended structure, apparently high in temperature, which we interpret as the observational signature of a current sheet. In particular, we present the first conclusive EUV evidence that features associated with current sheets can be visible to large heights in the corona, heights well above the edge AIA field of view, where similar features have been previously observed. Finally, we show the strong EUV wave that accompanied the eruption and that impacted the corona on a global scale. This wave was visible to the edges of SUVI's field of view at about 1.9~solar radii. 

Each of these phenomena has been relatively well observed, but only a handful of observations exist of these phenomena at heights as large as those reported here. For example, the trajectory of an eruption as it passes through the inner and middle corona can provide useful information about the nature of the acceleration mechanism of that eruption \citep{2008ApJ...674..586S}, but the coverage gap between AIA and coronagraphs like the Large Angle Spectroscopic Coronagraph \citep[LASCO;][]{1995SoPh..162..357B} on SOHO mean complete trajectories are limited to relatively few observations that include observations from large field of view instruments, such as the report in \citet{2017JSWSC...7A...7D}.

Likewise, there are various reports of observations of apparent current sheets in the EUV \citep[for example]{2017ApJ...835..139S, 2015SoPh..290.2211G, 2016ApJ...821L..29Z} and in rarer cases in both X-rays and visible light coronagraph observations as well \citep{2010ApJ...722..329S}. However, EUV and X-ray observations of these features have generally been limited by instrumental fields of view to a few tenths of a solar radius above the limb, and coronagraph observations limited to heights above about 2~solar radii. This has left a gap in observations at heights between about 1.4--2 solar-radii, where observations are needed to test a variety of important model predictions about the temperature \citep{2009ApJ...701..348S} and dynamics \citep{Forbes_CS_Submitted, 2012MNRAS.425.2824M} of reconnecting current sheets that are expected to form in the wake of an eruptive flare.

EUV waves --- sometimes referred to in the literature as ``EIT Waves'' --- have also been a subject of debate since their discovery about 20 years ago \citep{1997IAUJD..19E..18D, 1998GeoRL..25.2465T}. There is increasing consensus that these phenomena are fast-mode waves or shocks, driven by the impulsive expansion of a rapidly erupting CME \citep{2017SoPh..292....7L} but the nature of their interaction with the corona on larger scales, particularly at heights somewhat above the solar surface, remains a subject of vigorous research \citep[see][for just one example]{2013SoPh..286..201K}.

At the time of this writing, SUVI is still undergoing calibration and testing, and its data are generally still embargoed. However, given the genuine uniqueness of the observations we report here, we are making SUVI data from the event described in this paper available to the community immediately. We hope that providing rapid access to these data can aid researchers already working on analyses of the events discussed in this paper based on other observations. We believe that these novel observations of the extended EUV corona can help address many of the still-unanswered questions about the phenomena discussed above.

In Section~\ref{sec:inst} we discuss some key properties of the SUVI instrument and the data it produces in some additional detail. In Section~\ref{sec:erupt} we present SUVI's observations of the September 10 eruption and associated phenomena. In Section~\ref{sec:disc} we make some concluding remarks about the observations and what could be learned from them and how other researchers can access SUVI data for this event.

\section{SUVI Instrument and Data} \label{sec:inst}

SUVI is a new EUV solar telescope on board NOAA's new series of satellites in the Geostationary Operational Environmental Satellite (GOES) mission. GOES-16, the first of the GOES-R series of satellites, was successfully launched on 2016 November 19 and achieved geostationary orbit on 2016 November 29. The primary mission of SUVI --- and of the other of space weather instrumentation on board the GOES platform --- is to provide continuous in-situ observation of the near-Earth space environment and remote sensing observations of the Sun. Data from SUVI primarily support NOAA's capabilities to characterize solar features and detect events that might spawn space weather at Earth and nearby space environs.

As is the case for all GOES instrumentation, SUVI is designed to meet the operational needs of the National Weather Service's Space Weather Prediction Center (SWPC). To meet its operational requirements, SUVI has leveraged the successful solar imager designs of heritage instruments on past science missions, most notably AIA, with which it shares considerable design characteristics.

SUVI is a normal incidence EUV telescope in the Ritchey-Chr\'etien configuration with the detector at the Cassegrain focus. The use of filters and multilayer mirrors fine-tunes the six passbands in which SUVI images to correspond with known coronal EUV emission lines. These lines are 9.4~nm (Fe~\textsc{xviii}), 13.1~nm (Fe~\textsc{xxi}), 17.1~nm (Fe~\textsc{ix}/\textsc{x}), 19.5~nm (Fe~\textsc{xii}), 28.4~nm (Fe~\textsc{xv}), and 30.4~nm (He~\textsc{ii}). Figure~\ref{fig:onset} shows the onset of the eruption in each of SUVI's six bands, and more generally provides an overview of the SUVI field of view and appearance of the corona in each SUVI passband.

SUVI pixels are 2.5~arcsec in both x- and y-directions, meaning its 14-bit, $1280 \times 1280$ CCD detector has a total field of view of about 53.3~arcmin on the horizontal. SUVI's field of view extends farther in the diagonal direction, but vignetting in one or two corners of each passband limits the field of view slightly in some diagonal directions. Each passband has at least two corners that are unaffected by vignetting. This allows SUVI to image the corona to heights above 1.6~solar-radii on the horizontal and heights as large as 2.3~solar-radii on the diagonal.

An instrument paper describing the SUVI instrument in detail is currently in preparation and will be released in early 2018, coinciding roughly with routine SUVI data becoming publicly available. This paper will describe in detail aspects of the instrument hardware, the ground processing to Level-1b, Level-2 products of interest, and methods for accessing the data \citep{Darnel_SUVI}.

SUVI observes the EUV corona with a cadence of 10~s using an observing sequence that allows the imager to capture at least one image in every passband per four-minute observation cycle. The instrument automatically acquires calibration observations, using two of the 24 synoptic observation slots in each four-minute cycle, using on board logic that determines when new calibrations are necessary. At the time of the event described in this paper, routine observations were only interrupted, at most, for two frames out of every 720 observations, or once every two hours.

SUVI also generally obtains observations in sets of two or three in a given passband, using multiple exposures and filter combinations to increase the observable dynamic range by two or three orders of magnitude, depending on the passband. These image sets can then be combined by software on the ground to generate high-dynamic-range composites that do not include saturated pixels for flares as bright as X10. (In section~\ref{sec:erupt} we will describe how we used this compositing technique to image the bright flare core in the 13.1~nm passband.)

Because the SUVI instrument was not fully operational at the time of these observations, the data presented here were calibrated manually, using custom calibration codes written in {\sf IDL}. These codes draw on algorithms from the {\sf SolarSoft} software packages, including several used for calibrating SWAP data. (SWAP-derived calibration is especially appropriate for SUVI given the two instruments' similarities.) Some SUVI analysis and calibration codes are already available in {\sf SolarSoft}, although we expect to improve and extend this software library considerably once SUVI's post-launch testing period is complete and routine data are released.

Our calibration code ingests raw SUVI data, subtracts the appropriate bias and dark frames, corrects for a small amount of nonlinearity in the detector, corrects the flat field (uniquely for each passband and filter combination), performs some despiking to remove bright pixels from energetic particle hits, and converts the data to radiometric units. The code outputs FITS files in which spacecraft and guide-telescope auxiliary data are used to update metadata to provide accurate pointing and image orientation information.

We further corrected these observations by co-aligning frames to account for some slight jitter ($\lesssim0.3$~pixels, RMS) in the telescope pointing. The SUVI observations discussed subsequently in this paper come from these purpose-made files.

\section{The 2017 September 10 Eruption} \label{sec:erupt}

On 2017 September 10 at approximately 15:40~UT, SUVI observed the onset of a large eruption from the Sun's west limb. The eruption began with a rapid brightening of a small region near the limb in several SUVI channels, which was accompanied by the rapid rise of a flux rope and formation of a CME. 

The GOES X-ray irradiance curve for this event rises in two phases. The event begins with a jump from background levels that lasted just a few minutes, coincident with the initial onset of the event in SUVI images. The X-ray curve levels off for about five minutes, then begins a steady rise to the X8.2 level, which is roughly coincident with the rapid acceleration of the flux rope itself.

SUVI observations of the flare suggest that the footpoints of the flaring active region were at least partially occulted behind the solar limb, meaning there is a reasonable chance the X8.2 X-ray peak is an underestimate. In fact, there is evidence a significant fraction of a flare's X-ray flux can come from the flare footpoints \citep{2007A&A...461..315T}, although there is no simple way to determine the distribution of X-ray flux over the flaring region for this event due to its location. Nonetheless, since peak X-ray flux is sometimes used as a proxy for the energetics of an eruption more generally, it is worth noting this one is surely among the most energetic events of the present solar cycle --- if not the most energetic event.

\begin{figure}
\centering
\includegraphics[width=7in, keepaspectratio]{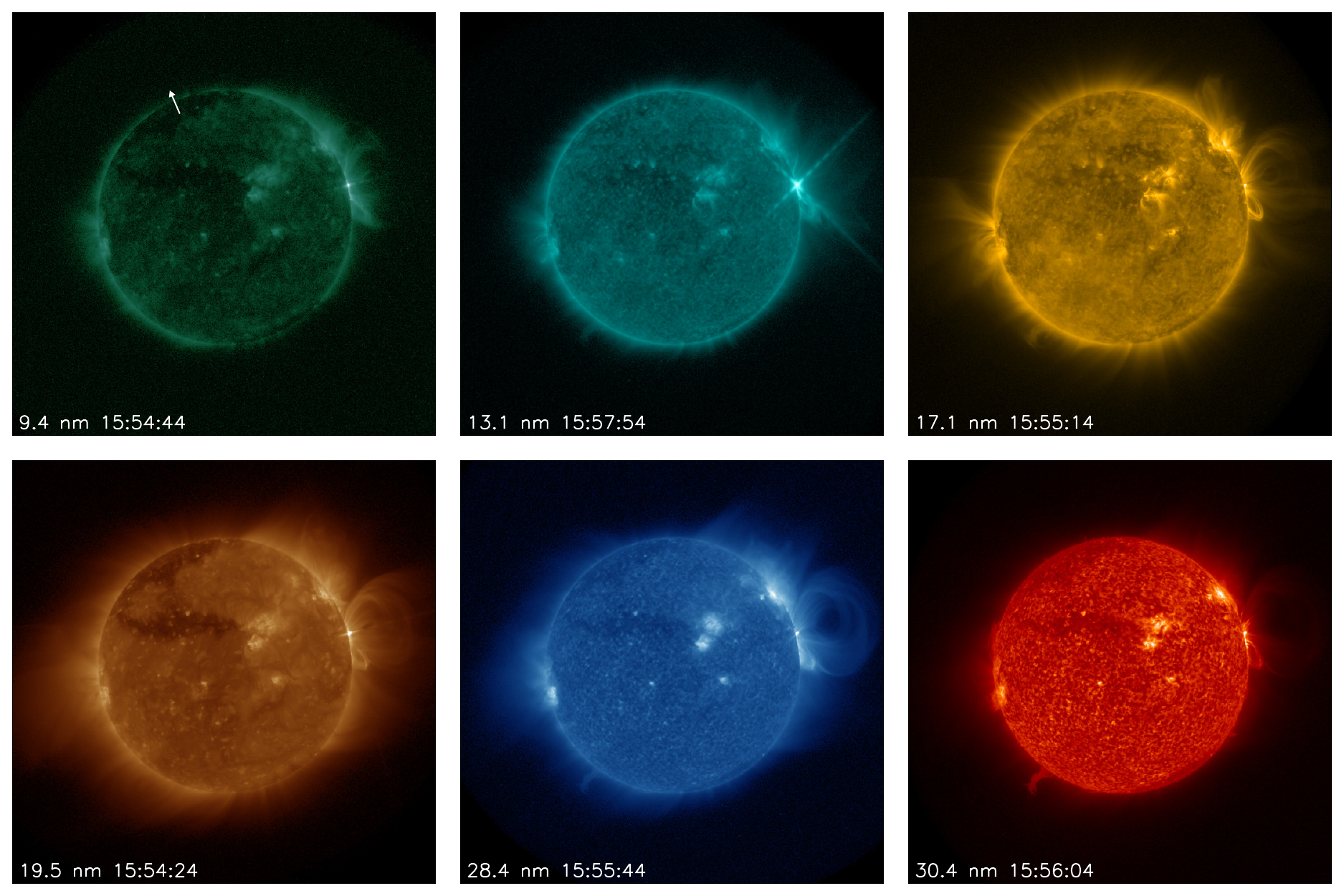}
\caption{Overview of SUVI's view of the corona just after the onset of the event. Each image is a composite of long and short exposures to eliminate saturation, and is contrast-adjusted for maximum visibility of the corona on global scales. Because SUVI's camera is generally aligned to terrestrial north, an arrow overlaid on the 9.4~nm frame indicates the location of the Sun's pole and direction of solar north. Note that the 28.4 and 30.4~nm bandpasses are wide enough that there is some cross-contamination from the 30.4 line in the 28.4~nm image and vice versa.}
\label{fig:onset}
\end{figure}

Figure~\ref{fig:onset} shows an overview of SUVI's view of the corona just after the time of the onset of the event. Each of the six frames is a composite of long and short exposures to eliminate saturation. (Although, to achieve reasonable display of the corona on global scales we sometimes clipped the image dynamic range to exclude the very bright flare kernel.) The bright flare emission is most clearly visible on the west limb in the 13.1~nm frame, while images in cooler passbands --- especially 19.5 and 28.4~nm --- show the rapidly expanding flux rope as it is accelerated into a CME.

\subsection{CME Trajectory} \label{subsec:traj}

\begin{figure}
\centering
\includegraphics[width=7in, keepaspectratio]{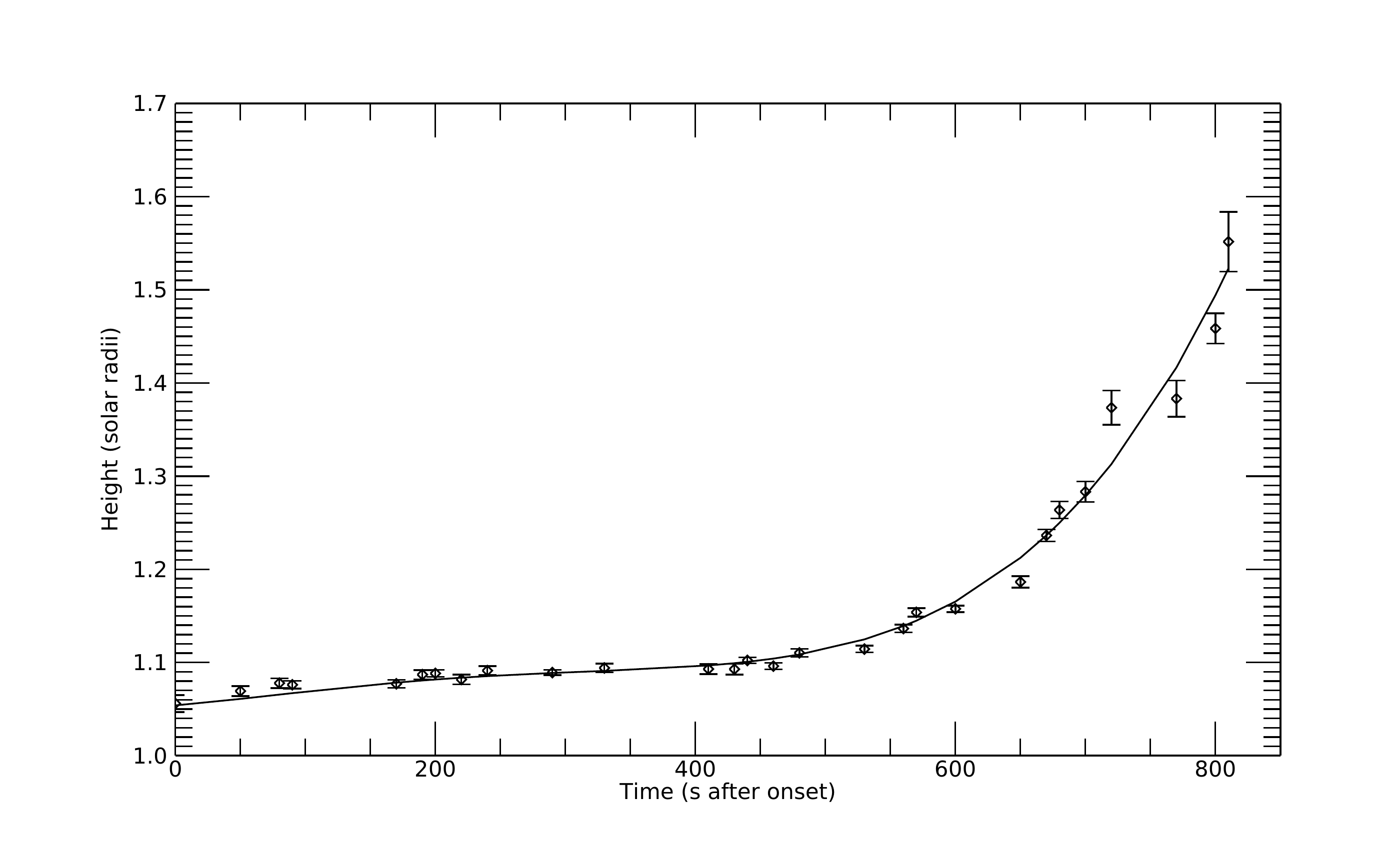}
\caption{Trajectory of the flux rope as it was accelerated through SUVI's field of view beginning at 15:44:04~UT. The smooth curve shows the fitted fourth-degree polynomial we used to estimate the velocity of the eruption as a function of time.}
\label{fig:rise}
\end{figure}

Beginning with an observation at 15:44:04~UT\footnote{All times referenced in this paper are in UT, but for the sake of brevity we omit this notation in the subsequent discussion}, we tracked the center of the rising flux rope in every SUVI image in which we could locate it until it approached the edge of SUVI's field of view after roughly 800~s. Figure~\ref{fig:rise} shows the trajectory we measured as a function of time. We fit the measured trajectory with a fourth-degree polynomial to provide a smoothly varying height-time profile from which we could estimate the velocity of the flux rope as a function of time. Note that because of this eruption's location very close to the limb, projection effects are negligible.

According to our estimates, the flux rope rose slowly until about 500~s after our first measurement, which itself was several minutes after the onset of the event. At this point the flux rope was accelerated much more rapidly. Using our fitted curve, we estimate the velocity of the flux rope at the last time we could accurately locate its center, at a height of about 1.55~solar radii, was around 2000~km/s. This is a very high velocity for a CME \citep{2005ApJ...619..599Y}, especially in light of the fact that many CMEs experience considerable acceleration at heights well above this \citep{2011ApJ...738..191B}, which is another indication of just how energetic this event was.

\subsection{Current Sheet \& Event Initiation Mechanism} \label{subsec:cs}

\begin{figure}
\centering
\includegraphics[width=7in, keepaspectratio]{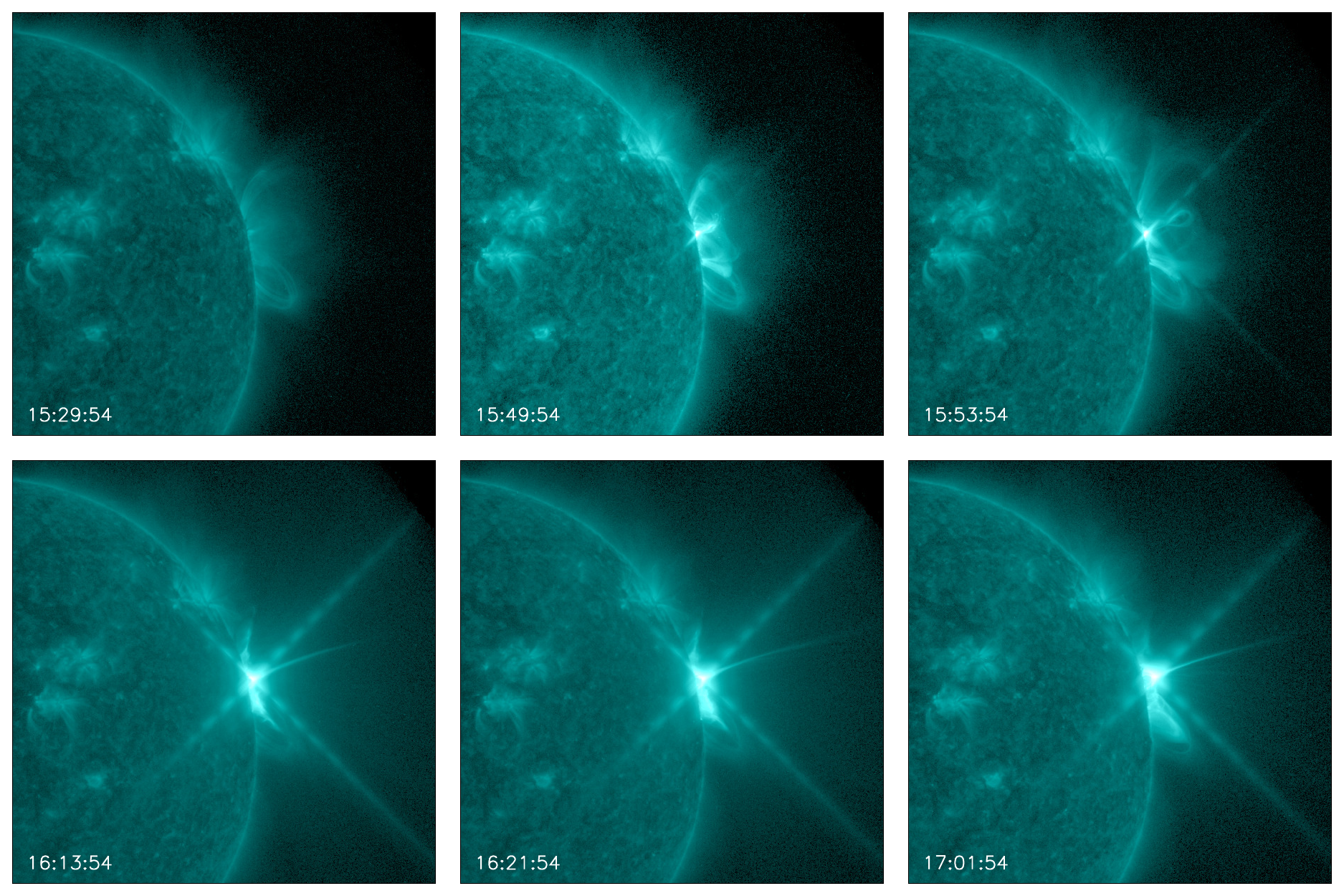}
\caption{Selected observations of the eruption and flare region in SUVI's 13.1~nm passband. The images are log-scaled. The upper left frame (15:29:54) shows the region just before the onset of the event. Subsequent frames in the top row show the early evolution of the eruption, and frames in the second row (16:13:54--17:01:54) show the narrow structure we interpret as a current sheet and its evolution over the course of the event. Refer to the accompanying online animation to see the complete evolution of the region in this passband. This animation is available at \url{https://data.ngdc.noaa.gov/platforms/solar-space-observing-satellites/goes/goes16/l1b/suvi-l1b-fe131/event_movies/2017/09/10/NCEI_SUVI_131_G16_HDR_20170910.mp4}.}
\label{fig:movie131}
\end{figure}

Figure~\ref{fig:movie131} shows the evolution of the region in SUVI's 13.1~nm passband. Observations of flaring regions in this passband are generally dominated by emission from Fe~\textsc{xxi} \citep{2010A&A...521A..21O}, which has a peak temperature just over 11~MK. Some of the background elsewhere in the corona and pre-flare structures in the active region of interest likely correspond to a cooler line in the passband, Fe~\textsc{viii}, which has a temperature closer to 0.4~MK.

The images in the figure and the accompanying animation are composites of two or three SUVI frames. Most of the corona is captured using an exposure of approximately 1~s, while a second exposure of 5~ms captures the brightest features, which will generally saturate in the longer exposure. A third exposure, also 5~ms, adds a second filter to the optical path, further reducing the incoming flux and allowing SUVI to clearly image pixels with radiances up to about $2\times10^{4}\,\mathrm{W\,m^{-2}\,sr^{-1}}$ in this band. For each frame of the figure and animation, these three frames have been composited by software on the ground that detects saturated pixels and replaces them with their unsaturated counterparts from one of the shorter exposures. If no pixels are saturated in the long exposure, which has the best signal-to-noise ratio, the short exposures are omitted from the composite. The images here are displayed on a log scale because of the large dynamic range of the flaring region.

Early in the evolution of the event (see the frames at 15:49:54 and 15:53:54) we see the apparently hot, rising core of the erupting flux rope situated between two more diffuse lobes that appear to retract as the event unfolds. Later, we see evidence of strong heating near the footpoints of the lobes, suggesting these features may have been heated by magnetic reconnection that must have occurred above the erupting flux rope.

As the flux rope continues to rise, a long, narrow spine-like structure forms behind it, eventually extending out to the edge of SUVI's field of view (see the frame at 16:21:54). At its furthest extent, this structure is visible to a height of at least 1.67~solar radii, which is the highest SUVI can image in the direction the structure extends. The movie accompanying Figure~\ref{fig:movie131} shows clear evidence of flows and dynamics along this structure, including several outflows. (One clearly occurs between about 16:45 and 17:15.)

This structure bears a strong resemblance to AIA observations of a current sheet associated with an X4.9 solar flare that occurred on 2014~February~25 reported by \citet{2017ApJ...835..139S}. Differential Emission Measure (DEM) reconstructions of that structure revealed a peak temperature around 10~MK, and running difference movies revealed flows along --- and into --- the sheet. However, those observations only allowed the characterization of the feature at relatively low heights in the corona, due to the limitations of AIA's field of view. Here we show clear evidence that current sheets such as these can extend to considerably larger heights.

It is worth pointing out that this structure is also clearly observed in other SUVI passbands. A complete analysis of the structure and its properties is beyond the scope of this Letter, but it is evident that SUVI will be a useful new tool for probing the properties of features related to magnetic reconnection in the higher EUV corona. Such observations can help better constrain a number of model predictions about the nature of eruption-associated reconnection.

\subsection{EUV Wave} \label{subsec:wave}

\begin{figure}
\centering
\includegraphics[width=7in, keepaspectratio]{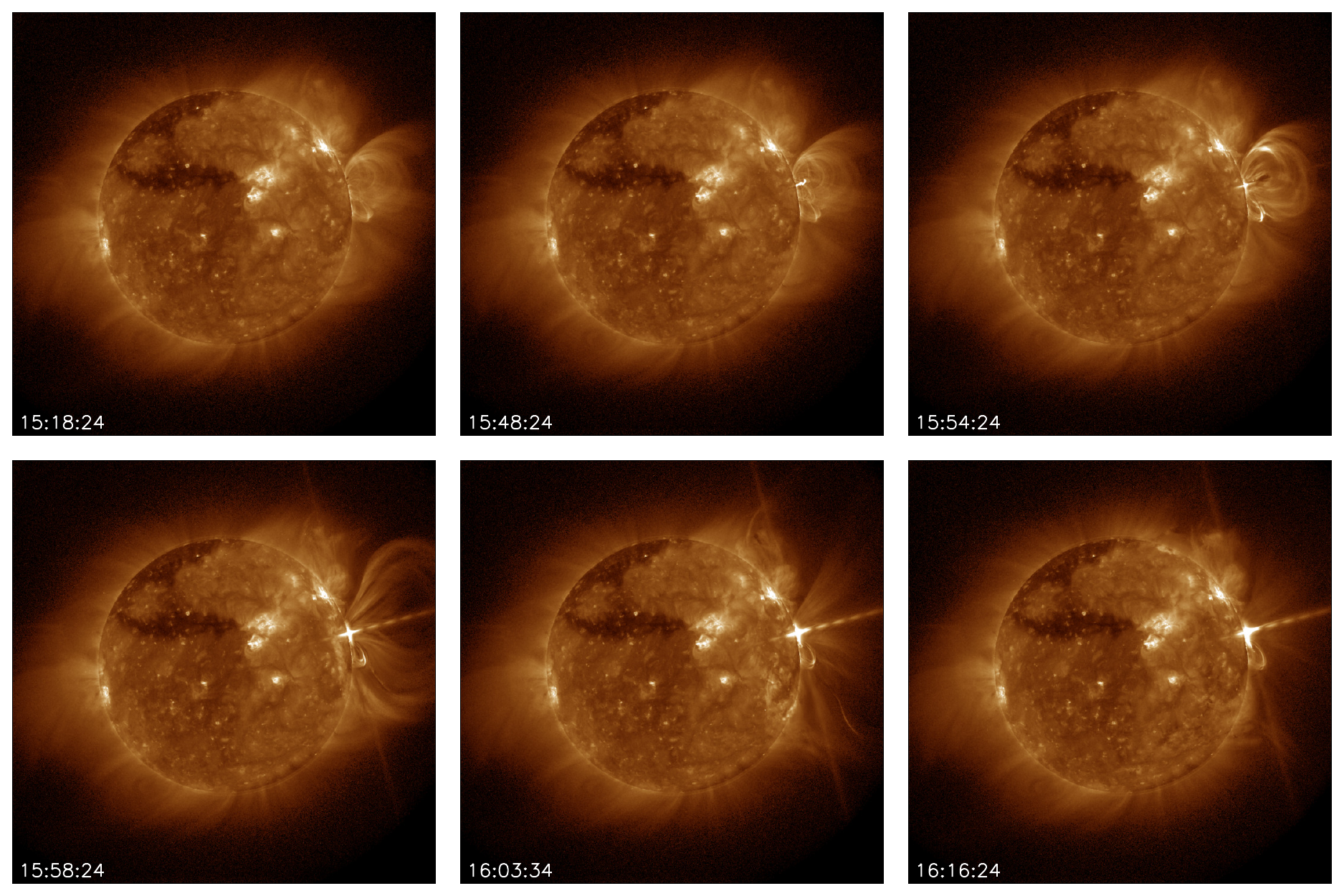}
\caption{Selected observations of the eruption, flare, and associated EUV wave in SUVI's 19.5~nm passband. These images have been contrast normalized and processed with a filter that enhances the contrast of dynamic events. The upper left frame (15:18:24) shows the corona just before the onset of the event. Subsequent frames in the top row show the early evolution of the eruption, and frames in the second row (15:58:24--16:16:24) show the propagation of the EUV wave across the entire solar disk. Refer to the accompanying online animation to see the complete evolution of the region in this passband. This animation is available at \url{https://data.ngdc.noaa.gov/platforms/solar-space-observing-satellites/goes/goes16/l1b/suvi-l1b-fe195/event_movies/2017/09/10/NCEI_SUVI_195_G16_DYNAMIC_20170910.mp4}.}
\label{fig:movie195}
\end{figure}

Figure~\ref{fig:movie195} and the associated animation show the evolution of the event including the early rise of the flux rope and propagation of the associated EUV wave across the Sun. 

These images have been processed with a two-step filter that enhances the visibility of dynamic structures in the observation. The first step is an azimuthally varying radial filter that reduces the total dynamic range of the corona as it falls off with height. This filter is not a true normalizing filter, but is tuned so that it neither amplifies noisy pixels at the edge of the image frame nor completely suppresses the gradient in coronal brightness with height. Thus it reduces the dynamic range of the corona enough that small scale structures are clearly visible in the images, but not so much that the filtered image appears unnatural.

In spite of this filter, the EUV wave in particular is still very faint compared to the overall brightness of the corona. To improve its visibility we apply a second filter that amplifies time-varying signals in each pixel of the image sequence. This filter is essentially the temporal analog of the well known photographic filter technique known as ``unsharp masking''. Fast-varying signals are detected by computing the difference between the each pixel and a seven-frame running median of the value of that pixel centered on the time of interest. The result is then amplified by an empirically determined factor, optimized for best display, and added back to the base image. Pixels that experience little variation retain their intrinsic brightness, while pixels that are experiencing a rapid variation appear amplified in brightness. Dynamic features are clearly visible in an image that nonetheless retains its natural appearance and avoids the artifacts induced by techniques such as running-difference imaging.

These processed images help reveal two important features of the eruption. First, we clearly see the impulsive acceleration of the flux rope that eventually became the core of the CME. The rising flux rope is especially clearly seen in the figure in the frames from 15:54:24 and 15:58:24. The kinematics of the CME have been discussed in detail in Section~\ref{subsec:traj}.

Second, we clearly see the EUV wave associated with the event. The wave first becomes clearly visible above the solar limb around the time the flux rope begins its impulsive acceleration and expansion. The wave quickly expands, reaching the far side of the solar disk within about 35~minutes. Assuming the wave traveled more or less directly across the surface of the sun, this means the wave must have propagated with a velocity of at least 1000~km/s. Waves with such fast speeds have been observed previously, but this velocity certainly places the wave among the fastest ever detected \citep{2017SoPh..292....7L}.

The wave also experiences several important interactions with the corona as it travels across the solar disk. For example, the wave appears to be reflected off the boundaries of the polar coronal holes. This reflection is apparent in the movie in the frames around starting 16:08 for the southern hole, and about ten minutes later for the northern hole.

Likewise, the wave disrupts considerably the bright region on (and above) the solar limb to the north of the active region that produced the eruption. The wave deforms the magnetic structures in this region as it passes through beginning around 16:00, inducing an oscillation that continues for at least 2.5~hours. As the wave passes through this region, an area of enhanced brightness is visible out to heights of at least 1.9~solar radii. Typical EUV waves are observed roughly inside the density scale height of the corona \citep{2017SoPh..292....7L} around 90~Mm, or roughly 1.15~solar radii.

These observations  represent one the first detections of the signature of an EUV wave at such large heights. Observations at these large heights could help shed light on the nature of EUV waves themselves. The appearance of an EUV wave at almost 2~solar radii supports the interpretation of \citet{2008SoPh..247..123D} that EUV waves are not low-altitude phenomena, but rather are high-altitude phenomena that have generally not been observed at large heights due to observational constraints.

Indeed, we tend to agree with this interpretation. It seems unlikely that we have detected the first EUV waves at large heights because this event is so unusual --- although it is among the most powerful eruptions of the solar cycle --- but rather because SUVI has provided the first ever opportunity to do so. We suspect that future observations of eruptions with SUVI are likely to yield similarly new observations thanks to the unique characteristics of the instrument.

\section{Discussion} \label{sec:disc}

In this Letter we have demonstrated the unique utility of SUVI for the observation of EUV coronal phenomena at large heights, particularly for eruptions. We have presented one of the first clear detections of an EUV wave at large heights. Likewise, we have presented a clear detection of high-temperature plasma associated with a current sheet undergoing magnetic reconnection due to an eruptive flare at heights above 1.5~solar-radii in the EUV.

Other authors \citep{2010ApJ...722..329S} have also reported reconnection-associated phenomena in X-rays at similar heights when projection effects are taken into account (and much larger heights in coronagraph observations). In terms of physical height above the solar surface, these SUVI observations are only one of several. On the other hand, these SUVI observations are to our knowledge the first at such heights in the plane of the sky, and they represent a first step towards filling the key observational gap between other EUV and coronagraph observations.

The observations that SUVI will make, like those presented in this paper, are especially important in light of the fact that they can help place key constraints on model predictions. For example, analytical models of reconnection during an eruptive flare by \citet{Forbes_CS_Submitted} suggest that the flow stagnation point in a reconnecting current sheet --- that is, the dividing point between outgoing reconnected flux and inflowing flux --- should be relatively low in the corona. To date, few observations have revealed the location of this point, which is essential for understanding the reconnection rate and the energy partition between a flare and CME during an eruption.

Although more analysis will be necessary to confirm the location of the stagnation point, we see evidence of outflows in along the current sheet in our observations, from which we might confirm that the model prediction by \citeauthor{Forbes_CS_Submitted} could be correct. Such a confirmation would help further constrain these models, and could have implications for the heating of plasma around the current sheet and in the outflow region near the top of the flare arcade, as discussed by \citet{2009ApJ...701..348S}.

These predictions aim to answer fundamental questions about the nature of solar eruptions that, eventually, could help us model and predict such events more accurately. SUVI can help answer these important questions. In doing so it will not only advance science in general and simultaneously carry out its operational mission for SWPC, but provide added value by improving SWPC's ability to interpret SUVI's observations themselves.

To help facilitate the research that can achieve this synergy, we will make our data available to the solar physics community, prior to the official release of SUVI files, both from this event and from the large on-disk eruption and flare that occurred a few days before on 2017 September 6. Users can find links to these SUVI files in FITS format and documentation at \url{https://www.ngdc.noaa.gov/metaview/page?xml=NOAA/NESDIS/NGDC/STP/Space_Weather/iso/xml/suvi-l1b-goesr.xml&view=getDataView&header=none}. Images from the entire mission should be available via the same website as well as through {\it Virtual Solar Observatory} and {\it Helioviewer} sometime in the first half of 2018.

It is worth noting these data are still considered to be in ``beta'' status, and might not be appropriate for all purposes. Our team welcomes the feedback and questions of the community regarding SUVI and these data specifically. Queries can be addressed directly to the authors of this paper.

With three more flight models slated for launch in the coming years, SUVI is expected to be active for the next 20 years, meaning its observations will extend our record of synoptic coronal observations nearly through two complete solar cycles. In this paper we have demonstrated its complementary value alongside other imagers like AIA, enhancing our understanding of phenomena that could not otherwise be easily observed. We hope these observations are just the first of many successful opportunities for SUVI to help extend our knowledge of coronal physics to larger heights.

\acknowledgments

We thank the anonymous referee for providing many thoughtful comments that improved this paper. This analysis and the preparation of the SUVI data discussed in this paper were supported by was supported by the NOAA National Centers for Environmental Information through NA15OAR4320137. SUVI was designed and built at Lockheed-Martin's Advanced Technology Center in Palo Alto, California. We thank the LMATC SUVI team for considerable support during SUVI's post launch testing period that made these observations possible. In particular we recognize the efforts of Chris Edwards, Dnyanesh Mathur, David Sabolish, Ralph Seguin, Margaret Shaw, Lawrence Shing, Greg Slater, and Gopal Vasudevan. We also thank the GOES-R program office for assistance in making SUVI data for these events available to the scientific community. We thank Steven Hill at SWPC for helpful conversations through SUVI's development and testing period. We thank our colleague Margaret Tilton for her assistance in preparing for the distribution of SUVI data files. 

\vspace{5mm}
\facility{GOES}

%% \software{SolarSoft, IDL, SunPy, Python}

%% The reference list follows the main body and any appendices.
%% Use LaTeX's thebibliography environment to mark up your reference list.
%% Note \begin{thebibliography} is followed by an empty set of
%% curly braces.  If you forget this, LaTeX will generate the error
%% "Perhaps a missing \item?".
%%
%% thebibliography produces citations in the text using \bibitem-\cite
%% cross-referencing. Each reference is preceded by a
%% \bibitem command that defines in curly braces the KEY that corresponds
%% to the KEY in the \cite commands (see the first section above).
%% Make sure that you provide a unique KEY for every \bibitem or else the
%% paper will not LaTeX. The square brackets should contain
%% the citation text that LaTeX will insert in
%% place of the \cite commands.

%% We have used macros to produce journal name abbreviations.
%% \aastex provides a number of these for the more frequently-cited journals.
%% See the Author Guide for a list of them.

%% Note that the style of the \bibitem labels (in []) is slightly
%% different from previous examples.  The natbib system solves a host
%% of citation expression problems, but it is necessary to clearly
%% delimit the year from the author name used in the citation.
%% See the natbib documentation for more details and options.

\bibliographystyle{aasjournal.bst}

\end{document}